# Anisotropic Spin Polarization and magnetic spin hall effect in Ferromagnets

Jiabin Wang[1], Zhenhua Zhang[1,*], Wancheng Zhang[2], Jianxiong Zhao[1], Yong Liu[3], Rui Xiong[3], Zhihong Lu[1,*]

[1]School of Materials Science and Engineering, Wuhan University of Science and Technology, Wuhan 430081, People's Republic of China

[2]School of Electrical and Electronic Information Engineering, Hubei Polytechnic University, Huangshi 435003, People's Republic of China

[3]Key Laboratory of Artificial Micro- and Nano-structures of Ministry of Education, School of Physics and Technology, Wuhan University, Wuhan 430072, People's Republic of China

Contact author: zzhua@wust.edu.cn (Zhenhua Zhang) and zludavid@live.com (Zhihong Lu)

**ABSTRACT**: Spin-dependent transport in ferromagnets underpins the development of high-density spintronic memories. Spin-dependent transport in strong spin-orbit-coupled ferromagnets exhibits a significant anisotropy. Both the overall spin polarization during charge transport and the magnetic spin Hall conductivity are found to exhibit pronounced anisotropy when the magnetization is tilted away from the crystallographic easy axis or when the electric field is rotated relative to the crystal axes. These anisotropic responses originate primarily from spin-orbit coupling, which is identified as the key driver of the large anisotropy observed in ferromagnet. Furthermore, strain tunability of the magnetic spin Hall anisotropy is demonstrated, with tensile strain progressively enhancing the oscillatory amplitude of the spin Hall conductivity. These findings establish strong spin-orbit-coupled ferromagnets as a platform for anisotropic spin-current generation and field-free spintronic devices that exploit intrinsic material anisotropy for improved performance and energy efficiency.

# 1. Introduction

Spin-transport phenomena in magnetic systems are critical to spintronics, as they enable the generation of spin-polarized charge currents and order-parameter-tunable spin currents, which underpin a wide range of magnetic memory and logic devices **[1-3]**. Such spin currents can be generated via two distinct mechanisms: the conventional spin Hall effect (SHE), which is even under time reversal and arises from spin-orbit coupling **[4-6]**, and the magnetic spin Hall effect (MSHE), which is odd under time reversal and originates from the combined action of magnetism and spin-orbit coupling **[7-9]**. Calculations for bulk ferromagnetic metals bcc Fe, hcp Co, and fcc Ni reveal that the MSHE conductivity can be comparable in magnitude to the SHE, yet it exhibits a strong dependence on the electron lifetime and can even have the opposite sign **[10]**. Notably, because the MSHE is dominated by intraband (Fermi-surface) contributions, its magnitude scales inversely with the lifetime broadening, implying that in ultraclean samples the MSHE can far exceed the SHE **[8,10]**. These results demonstrate that the MSHE cannot be neglected in spin-orbit torque studies **[11,12]**. The MSHE allows a pure spin current to be generated with its flow and polarization directions both perpendicular to the electric field and parallel to each other, which is essential for field-free magnetic memory devices **[2]**.

Recent studies reveal that magnetic spin-transport phenomena in altermagnets and noncollinear antiferromagnets exhibit pronounced anisotropy **[13-16]**, which is intimately linked to the momentum-dependent anisotropic band splitting inherent to these two classes of materials. The physical origin of this anisotropy lies in their distinctive $\mathcal{T}$-odd character. In altermagnets, the staggered spin splitting in momentum space breaks time-reversal symmetry (TRS), giving rise to anisotropic spin-dependent transport **[16].** In noncollinear antiferromagnets, the time-reversal-odd spin Hall effect serves as the primary source of anisotropic spin currents **[14]**. However, the corresponding anisotropy in ferromagnetic materials has received little attention. It is known that the magnetic spin Hall effect (MSHE) originates from spin orbit coupling (SOC) and is related to the symmetry of the system **[4]**. Although SOC in ferromagnets gives rise to magnetic anisotropies including magnetocrystalline anisotropy and anisotropic magnetoresistance, it remains unclear whether the magnetization-dependent spin transport also displays such anisotropy in these strong SOC systems **[17,18]**.

Taking strongly spin orbit coupled FePt as an example, both theory and experiment have confirmed that this material exhibits significant uniaxial magnetic anisotropy **[19,20]**. Furthermore, first principles calculations by Zhang and coworkers show that when the magnetization direction is rotated from [001] (the easy axis) to [110], the anomalous Hall conductivity of FePt decreases by nearly a factor of two, and this anisotropy originates almost entirely from spin flip scattering processes **[21,22]**. This finding reveals a strong dependence of the transverse charge transport coefficient on the orientation of magnetization. However, in studies of spin-dependent transport, existing works typically address the influences of magnetization direction and electric field direction in isolation, without a unified analysis of their coupled effect **[4,23]**. It remains unclear whether the spin conductivity is unchanged when either the electric field or the magnetization direction is varied.

In this work, the coupled effects of magnetization and electric-field orientations on spin transport in the strong spin-orbit-coupled ferromagnet are systematically investigated. Both the overall spin polarization during charge transport and the magnetic spin Hall conductivity are found to exhibit pronounced anisotropy when the magnetization is tilted away from the crystallographic easy axis or when the electric field is rotated relative to the crystal axes. These anisotropies originate from the interplay of spin-orbit coupling, crystalline symmetry, and Fermi-surface topology. Strain tunability of the magnetic spin Hall anisotropy is further demonstrated. These findings establish strong spin-orbit-coupled ferromagnets as a platform for anisotropic spin-current generation and field-free spintronic devices that exploit intrinsic anisotropy for improved performance and energy efficiency.

## 2. Methodology

First-principles calculations were performed within the framework of density functional theory (DFT) using the Vienna ab initio Simulation Package (VASP) **[25-26]**. The ion-electron interaction was described by pseudopotentials, and the exchange-correlation functional was treated with the Perdew-Burke-Ernzerhof (PBE) parametrization of the generalized gradient approximation (GGA) **[27,28]**. A plane-wave basis set with a cutoff energy of 520 eV was adopted. The convergence criteria for total energy and residual forces during structural relaxation were set to $1 \times 10^{-5}$ eV and 0.01 eV/Å, respectively. For the structural optimization, the Brillouin zone was sampled using a Γ-centered $10 \times 10 \times 8$ Monkhorst-Pack *k*-point mesh. Self-consistent field (SCF) calculations were carried out with a $15 \times 15 \times 12$ Monkhorst-Pack grid to ensure accurate electronic structure determination. The spin Hall conductivity was computed with the WANNIER-LINEAR-RESPONSE code **[29]**, as described in Sec. (i) of the Supplemental Material **[30]**.

## 3. Discussion

In conventional collinear antiferromagnets, the two sublattice moments are strictly antiparallel, yielding a vanishing net magnetization **[31,32]**. The translation-time-reversal or parity-time symmetry preserves Kramers degeneracy, leaving the spin-up and spin-down bands degenerate without spin splitting [see **Fig. 1(a)**]. Altermagnets combine the large spin splitting of ferromagnets with the vanishing net magnetization of antiferromagnets **[33,34]**. In their collinear antiferromagnetic order, the two sublattices are connected by rotational symmetry rather than by translational or inversion symmetry, which induces nonrelativistic spin splitting in momentum space and yields alternating spin-polarized Fermi surfaces [**Fig. 1(b)**]. This spin-splitting effect allows a charge current to generate a pure spin current without strong spin-orbit coupling, with the spin polarization locked to the Néel vector and electrically tunable.

In ferromagnets, atomic magnetic moments are aligned in parallel, giving rise to a macroscopic net magnetization and significant stray fields. This macroscopic magnetization breaks time-reversal symmetry, leading to a rigid Zeeman-type splitting between the spin-up and spin-down bands. In conventional ferromagnets with high crystalline symmetry, the Fermi surface in momentum space is typically isotropic. However, for anisotropic ferromagnets, spin-orbit coupling intrinsically breaks the rotational invariance of the electronic band structure, rendering the Fermi surface anisotropic in theory. This momentum-space anisotropy directly manifests in the spin-dependent transport properties, leading to orientation-dependent spin diffusion lengths, spin relaxation rates, and spin-torque efficiencies. Thus, the degree of Fermi-surface deformation serves as a key indicator of the underlying spin-orbit coupling strength and provides a viable route for engineering anisotropic spin transport in ferromagnetic systems. In addition, magnetic anisotropy enables the Fermi surface to be modulated by changes in the magnetization direction, resulting in its compression and distortion [**Fig. 1(c)**].

Given that both spin-dependent transport and magnetic anisotropy originate from the spin-orbit coupling in the system, the anisotropic behavior of spin transport in magnetically anisotropic materials is systematically investigated in this work. The effects of the relative orientation between the electric field and the magnetization on the spin-polarized current and the magnetic spin Hall effect are particularly studied. Specifically, two scenarios are considered: varying the electric-field direction for a fixed magnetization, and varying the magnetization direction for a fixed electric field.

Ferromagnetic materials are adopted as a model system, for which the anisotropic spin polarization and magnetic spin Hall effect are calculated. As illustrated in **FIG. 1(d)**, with the magnetization oriented along [001], a spin-polarized current is generated along the electric field direction, serving as a probe for the anisotropy of spin polarization. To treat general magnetization directions, a local coordinate system is introduced with the $x_0$ axis along the magnetization, and an electric field is applied along $x_0$. The magnetic spin Hall conductivity $\sigma_{z_0x_0}^{z_0}$ is then computed in this frame [**Fig. 1(e)**]. Furthermore, by projecting the coordinate system onto the $y_0z_0$ plane and rotating the electric-field direction while the magnetization is kept fixed, the evolution of this conductivity with the field orientation is investigated [**Fig. 1(f)**].

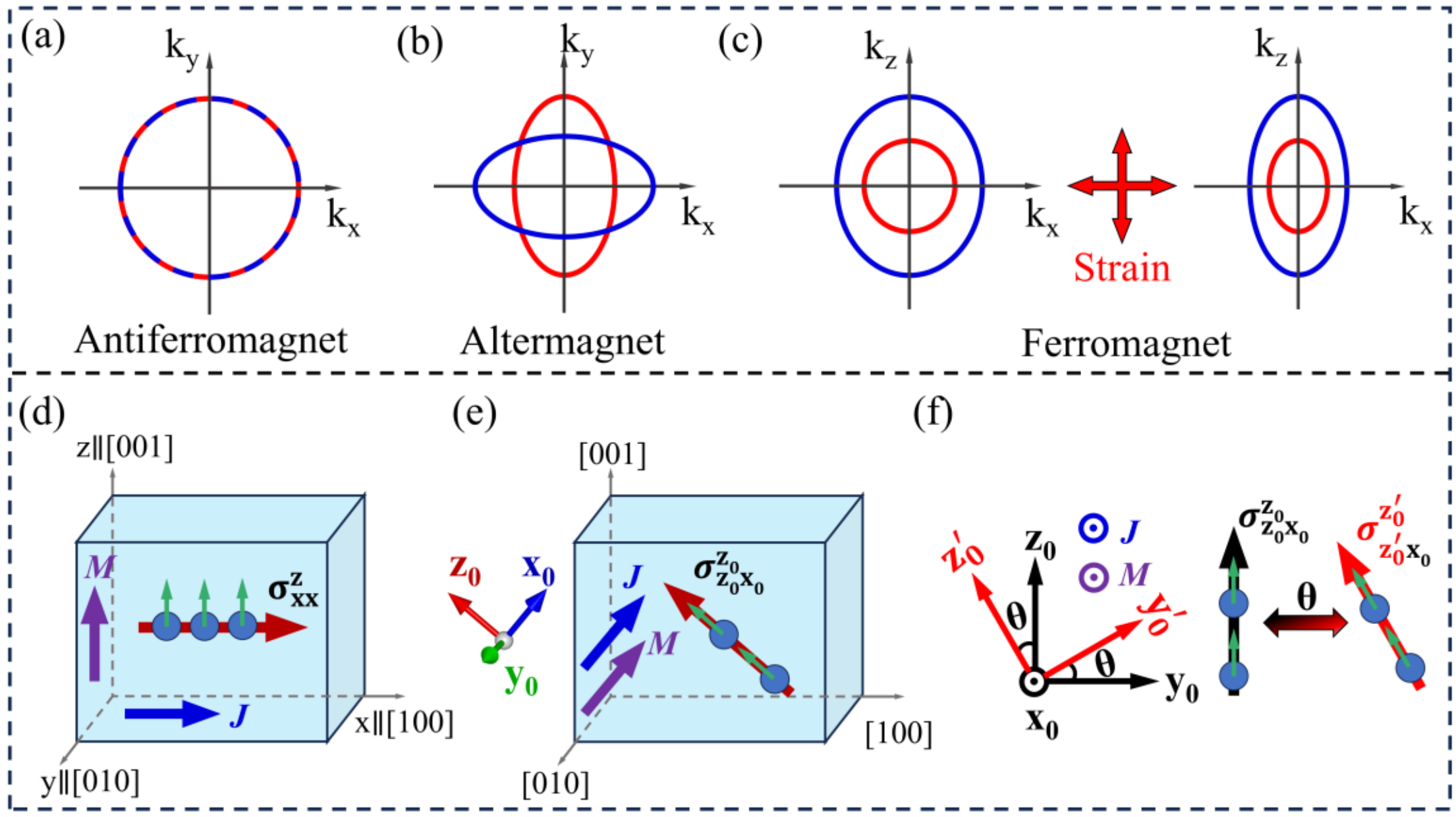

**FIG. 1.** Schematic of magnetic order and band structures for different material classes. (a) Collinear antiferromagnets, (b) altermagnets, (c) ferromagnets. Red and blue denote spin-down and spin-up Fermi surfaces, respectively. (d) Configuration for studying anisotropic spin polarization: magnetization along [001], electric field applied parallel

to the spin-current direction. (e) Local coordinate system with $x_0$ along an arbitrary magnetization direction, used to compute the magnetic spin Hall conductivity $\sigma_{z_0 x_0}^{z_0}$. (f) Angular evolution of $\sigma_{z_0 x_0}^{z_0}$ as the electric-field direction is rotated in the $y_0 z_0$ plane while the magnetization is kept fixed. The upper three-dimensional panels depict the magnetization orientation $M$ (purple arrow), the charge current $J$ (blue arrow), the spin-polarization axis (green arrow), and the spin-current propagation direction (red arrow).

FePt, a prototypical ferromagnet with strong spin-orbit coupling (SOC), serves as an ideal model system for investigating anisotropic spin polarization and the magnetic spin Hall effect in ferromagnetic materials. Symmetry analysis, as detailed in **Table SI** of the Supplemental Material, reveals that when the magnetization is oriented along the [001] direction, the time-reversal-even ($\mathcal{T}$-even) tensor components satisfy the relations $\sigma_{zy}^{x} = -\sigma_{zx}^{y}$, $\sigma_{yz}^{x} = -\sigma_{xz}^{y}$, and $\sigma_{xy}^{z} = -\sigma_{yz}^{z}$; all these components vanish in the absence of SOC. For the time-reversal-odd ($\mathcal{T}$-odd) sector, the components obey $\sigma_{zy}^{y,odd} = \sigma_{zx}^{x,odd}$ and $\sigma_{zy}^{y,odd} = \sigma_{zx}^{x,odd}$, which also disappear when SOC is neglected. This confirms that both the conventional spin Hall effect and the magnetic spin Hall effect are of relativistic origin. Remarkably, however, the remaining $\mathcal{T}$-odd components $\sigma_{xx}^{z,odd}$, $\sigma_{yy}^{z,odd}$, and $\sigma_{zz}^{z,odd}$ persist even without SOC, indicating that they originate from magnetic exchange interactions rather than from relativistic effects. These observations, while not exclusive to FePt, reflect general symmetry constraints that apply to a broad class of magnetic systems. To investigate the angular dependence of the spin transport behavior, the anisotropic spin polarization in FePt is schematically illustrated in **FIG. 1(d)**, which depicts the underlying geometry and measurement configuration. The rotation of the spin conductivity and the charge conductivity about the [010] crystallographic axis is studied via symmetry analysis. Under this rotation, the spin conductivity components transform according to $\sigma_{i,j}^{s,k} = \sum_{l,m,n} D_{il} D_{jm} D_{kn} \sigma_{(001)l,m}^{s,n}$, while the charge-conductivity components follow $\sigma_{i,j} = \sum_{l,m} D_{il} D_{jm} \sigma_{(001)l,m}$. These transformation rules arise directly from the underlying rotational symmetry and provide a basis for interpreting the orientation dependence of the transport coefficients.

The spin polarization is equal to the ratio of the spin-diagonal conductivities to the charge conductivity, reaching its maximum at the in-plane configuration ($|\sigma_{\mathrm{xx}}^{\mathrm{z,odd}} / \sigma_{\mathrm{xx}}|$, $\psi = 0°$) and its minimum at the out-of-plane configuration ($|\sigma_{\mathrm{zz}}^{\mathrm{z,odd}} / \sigma_{\mathrm{zz}}|$, $\psi = 90°$). As shown in **FIG. 2(a)**, rotation of the electric field in the xz plane leads to a monotonic decrease in spin polarization with angle, with values at $|\sigma_{\mathrm{aa}}^{\mathrm{z,odd}} / \sigma_{\mathrm{aa}}|$ ($\psi = 30°$), $|\sigma_{\mathrm{bb}}^{\mathrm{z,odd}} / \sigma_{\mathrm{bb}}|$ ($\psi = 45°$), and $|\sigma_{\mathrm{cc}}^{\mathrm{z,odd}} / \sigma_{\mathrm{cc}}|$ ($\psi = 60°$) in descending order. This clearly establishes the anisotropic character of spin polarization in ferromagnetic materials. This observed trend underscores the sensitivity of the conversion process to the orientation of the electric field relative to the crystallographic axes, a feature expected to be of general relevance for understanding spin-transport phenomena in magnetic systems. Subsequently, the influence of the magnetization orientation on the overall spin polarization ($P = \sqrt{(\sigma_{\mathrm{zz}}^{\mathrm{x}})^2 + (\sigma_{\mathrm{zz}}^{\mathrm{y}})^2 + (\sigma_{\mathrm{zz}}^{\mathrm{z}})^2} / \sigma_{\mathrm{zz}}$) during charge transport along the z direction (i.e., the [001] crystallographic axis) is investigated. A clear dependence is observed: upon deviation of the magnetization from the [001] axis, the overall spin polarization (*P*) for charge transport along the z direction varies accordingly, as shown in **FIG. 2(b)**. The scattering rate is determined based on both the experimental resistivity of $L1_0$-FePt and the calculated electrical transport properties, yielding a value of $\Gamma$ = 100 meV adopted **[35]**. The computed conductivity of FePt at this scattering rate is presented in **FIG. S1** of the Supplemental Material; all subsequent analyses of the magnetic spin Hall conductivity are performed using the same value of $\Gamma$ = 100 meV. A magnified view of the high-scattering regime in **FIG. 2(b)** is provided in **FIG. S2**, where the overall spin polarization (*P*) is found to decrease monotonically with increasing magnetization tilt angle φ. These results indicate that the magnetization orientation, through its modulation of the electronic band structure and Fermi-surface topology, exerts a substantial influence on the overall spin polarization (*P*) in ferromagnetic systems. The observed anisotropic spin polarization, with its strong dependence on both the crystallographic direction and the magnetization orientation, offers practical pathways for spintronic device engineering.

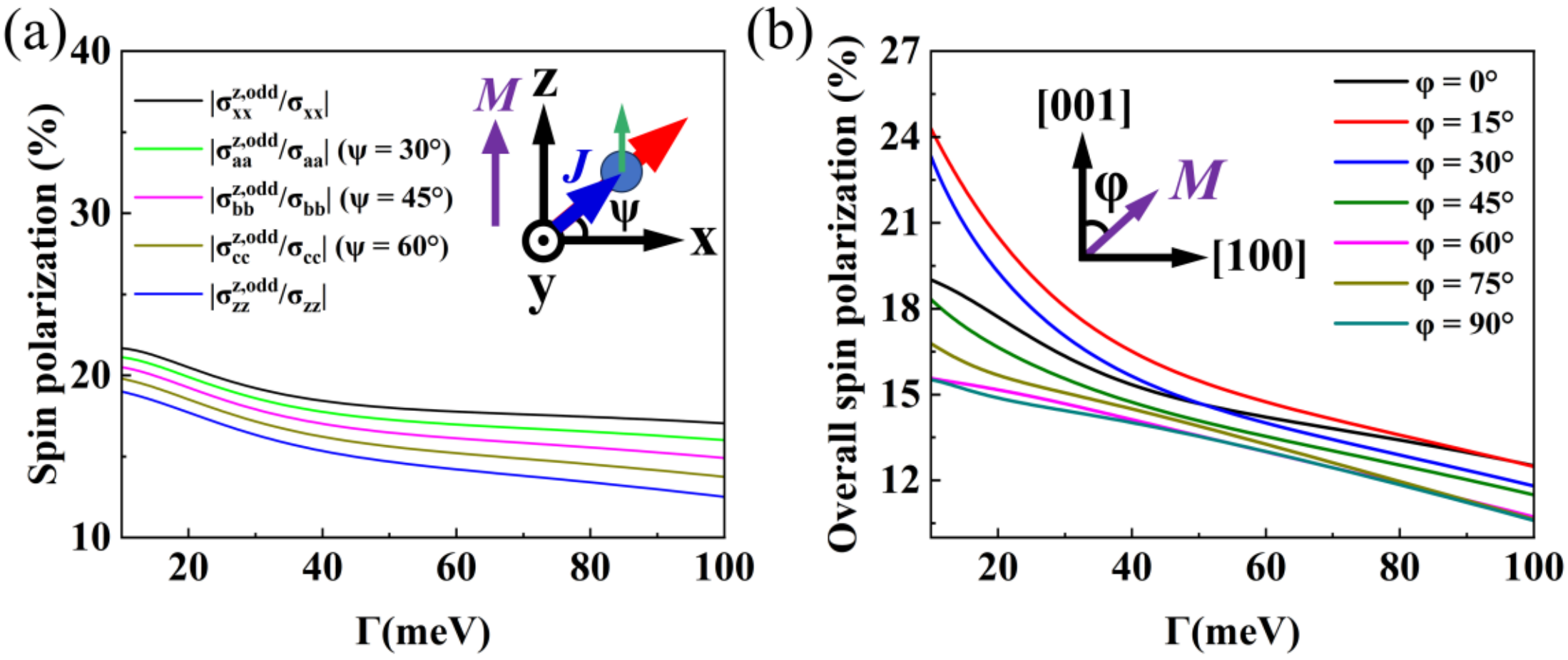

**FIG. 2.** (a) Spin polarization as a function of the spin-polarized current orientation in the xz plane for FePt with magnetization along [001]. (b) Overall spin polarization (*P*)

during charge transport along the z direction as a function of the magnetization tilt angle φ relative to the [001] axis for FePt. The inset illustrates the magnetization direction *M* (purple arrow), charge current *J* (blue arrow), spin-polarization axis (green arrow), and spin-polarized current propagation direction (red arrow).

While the foregoing results elucidate the pronounced anisotropy of spin polarization and its strong dependence on the magnetization orientation and current direction, the MSHE constitutes another key manifestation of spin-dependent transport that is also governed by crystalline symmetry and spin-orbit coupling. Next, the symmetry constraints and the angular response to both the magnetization and the applied electric field are addressed. Specifically, the behavior of the magnetic spin Hall conductivity components under symmetry operations and their variation with the relative alignment of the magnetization with respect to the crystal axes and the field direction are investigated. The matrix elements $\sigma_{\mathrm{yx}}^{\mathrm{y,odd}}$ and $\sigma_{\mathrm{zx}}^{\mathrm{z,odd}}$ are central to this study, since the out-of-plane spin current from the MSHE allows field-free switching of perpendicular magnetization when symmetry conditions are met, which is crucial for device miniaturization and low-power spintronic applications. As shown in **Tables SI-II** of the Supplemental Material, the magnetic spin Hall conductivity components $\sigma_{\mathrm{yx}}^{\mathrm{y,odd}}$ and $\sigma_{\mathrm{zx}}^{\mathrm{z,odd}}$ vanish for magnetization along [001] and become finite upon tilting away from this axis, indicating that the effect is symmetry-forbidden in the collinear [001] configuration. The interplay between the electric-field and magnetization orientations is systematically studied. In **Fig. 3(a)**, the magnetic spin Hall conductivity $\sigma_{zx}^{z,odd}$ (evaluated at a scattering rate Γ = 100 meV) is shown to vary with the magnetization direction for a fixed electric field. Conversely, for a given magnetization orientation, $\sigma_{zx}^{z,odd}$ exhibits a periodic oscillation as a function of the applied electric-field angle, with a 180-degree period. The oscillation amplitude increases with the tilt angle of the magnetization relative to [001]; it remains small at low angles but becomes pronounced at high angles. To further elucidate this behavior, the magnetization is fixed along [100] (φ = 90°), and the angular dependence of the conductivity is investigated under different scattering rates [**FIG. 3(b)**]. In all cases, the periodic modulation persists. The above results establish a pronounced anisotropy of the magnetic spin Hall effect in ferromagnets, with both the magnitude and the angular dependence of the conductivity found to depend strongly on the relative orientation of the magnetization to the crystal axes and the electric field. The 180° periodicity and the amplitude enhancement with tilt angle are ascribed to crystalline symmetry and spin-orbit coupling. The persistence of this behavior over a wide range of scattering rates indicates an intrinsic origin, suggesting a route toward electrically tunable spin-current generation in anisotropic magnets.

The microscopic origin of the observed anisotropies can be understood by examining the Fermi-surface topology, as schematically depicted in **FIGs. 1(c)** and **1(d)**. In FePt, the strong spin-orbit coupling intrinsically breaks the rotational symmetry of the electronic band structure, leading to a momentum-space Fermi surface that is appreciably deformed along different crystallographic directions. When the magnetization is rotated relative to the crystal axes, this deformation undergoes further compression and distortion, as illustrated in **FIG. 1(c)**, which in turn modulates the spin-dependent scattering and transport coefficients. Consequently, the orientation-dependent spin polarization and the magnetic spin Hall conductivity measured in our calculations directly reflect the underlying Fermi-surface anisotropy. This connection between the band-structure topology and the transport response provides a unified framework and reinforces the notion that magnetic anisotropy controls spin transport via Fermi-surface geometry.

To further substantiate the role of spin-orbit coupling in the observed anisotropic responses, complementary calculations were performed on $CrO_2$ (see **FIG. S3** in the Supplemental Material), a ferromagnet with uniaxial magnetic anisotropy but considerably weaker spin-orbit coupling than FePt. For $CrO_2$, the calculated anisotropy in the spin polarization is found to be substantially lower than that in FePt, despite the similar crystalline symmetry. This comparison confirms that the strong SOC in FePt is the primary driver of the large anisotropic effects observed here, and it underscores the importance of materials with heavy elements for achieving pronounced spin-transport anisotropy.

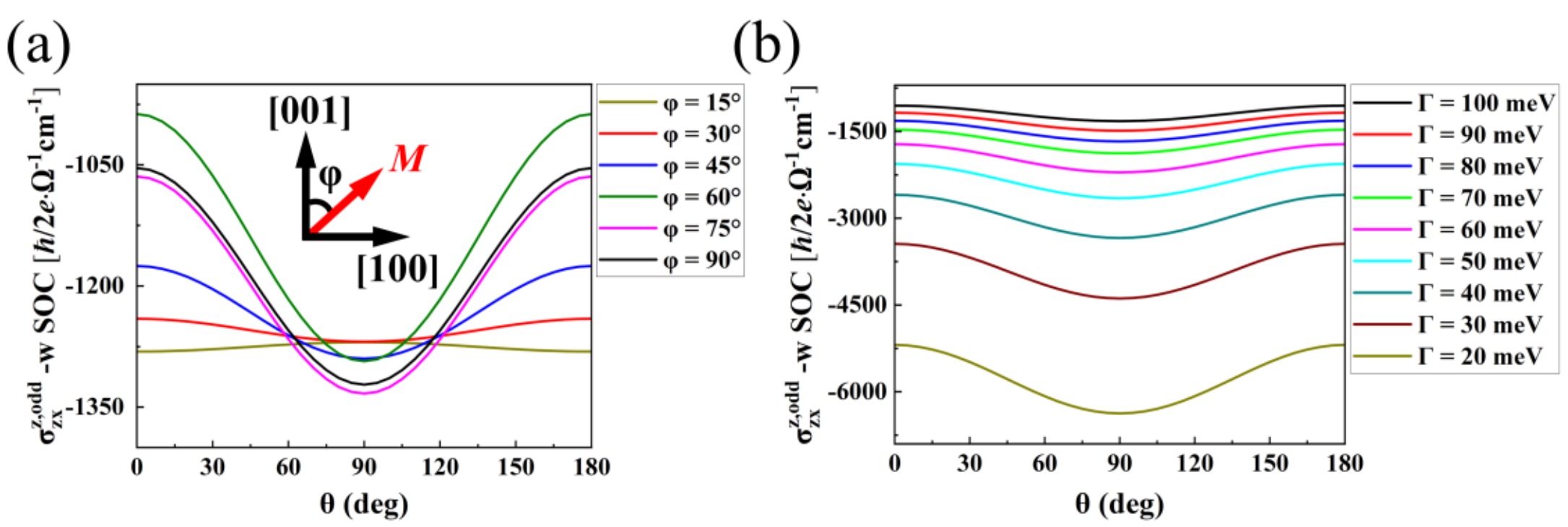

**FIG. 3.** (a) Magnetic spin Hall conductivity $\sigma_{zx}^{z,odd}$ as a function of the applied electric-field angle for various magnetization tilt angles relative to [001], evaluated at a scattering rate Γ = 100 meV. (b) Angular response of $\sigma_{zx}^{z,odd}$ to the electric-field direction at different scattering rates, with the magnetization along [100] (φ = 90°). The angle θ of the applied electric field is defined in Fig. 1(f).

The pronounced anisotropy observed in the spin-dependent transport properties of FePt is rooted in the underlying Fermi-surface topology, which is highly sensitive to the details of the electronic band structure. Since strain is known to effectively modify the band structure **[36,37]** and, consequently, the Fermi-surface anisotropy, it provides a versatile external knob for tuning the magnitude and angular response of the spin-Hall and charge-transport coefficients. To explore this possibility, the influence of tensile strain on both the overall charge transport along the z direction and the magnetic spin Hall effect is systematically studied, with the magnetization fixed along [100]. The overall charge transport along z increases monotonically with equibiaxial tensile strain, being minimal at zero strain and maximal at 4% strain [**FIG. 4(a)**]. At a fixed scattering rate of Γ = 100 meV, the magnetic spin Hall conductivity $\sigma_{zx}^{z,odd}$ under equibiaxial strain retains a clear 180°-periodic angular dependence on the electric-field direction [**Fig. 4(b)**]. Notably, the oscillation amplitude of $\sigma_{zx}^{z,odd}$ is progressively enhanced by the applied strain, with the smallest value at zero strain and the largest at 4% strain, indicating a high sensitivity of the anisotropic spin-Hall response to lattice deformation. Compared with the unstrained case, the magnetic spin Hall conductivity at 4% tensile strain approaches zero when the electric field is along θ = 0°, while remaining nearly unchanged at θ = 90°. These observations suggest that strain not only modulates the magnitude of the overall charge transport but also effectively amplifies the spin-Hall anisotropy, likely via strain-induced modifications of the band structure and spin-orbit coupling strength. The systematic evolution of both overall charge transport along the z direction and its angular response with strain underscores the potential of mechanical strain as a powerful external handle for controlling spin-current generation in ferromagnetic systems.

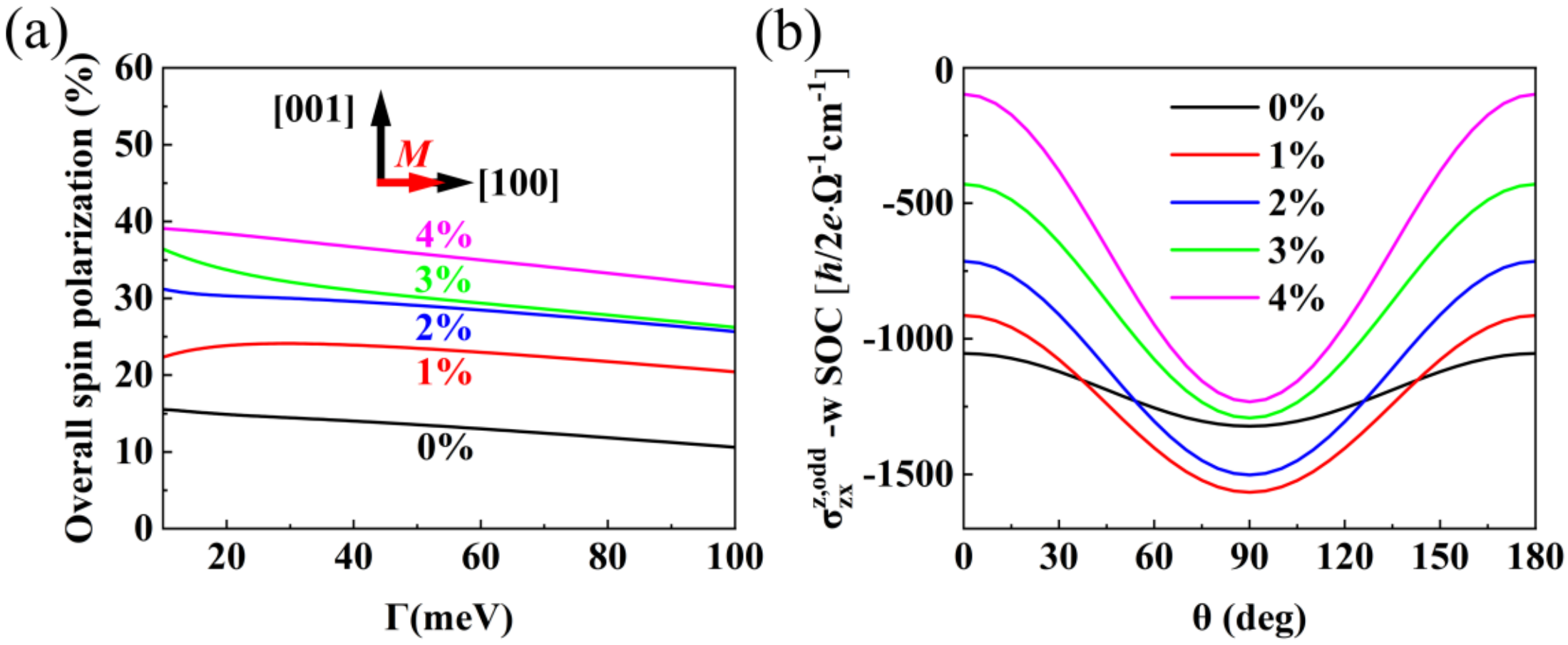

**FIG. 4.** (a) Overall charge transport along the z direction as a function of equibiaxial tensile strain, with the magnetization fixed along [100]. (b) Angular dependence of the magnetic spin Hall conductivity $\sigma_{zx}^{z,odd}$ on the electric-field direction under different strain values, evaluated at a scattering rate Γ = 100 meV.

## 4. Conclusion

In summary, the anisotropic spin polarization and magnetic spin Hall effect in strong spin-orbit-coupled ferromagnetic systems are systematically investigated, taking

FePt as a prototypical example. The results reveal a rich phenomenology governed by crystalline symmetry, spin-orbit coupling, and Fermi-surface topology. By comparing with the weakly spin-orbit-coupled ferromagnet $CrO_2$, the strong spin-orbit coupling in FePt is confirmed as the key factor responsible for the large observed anisotropy. The observed anisotropies and their sensitivity to the relative orientation of the magnetization, electric field, and crystal axes offer multiple pathways for device applications, including field-free spin-torque switching and electrically tunable spin-current generation. Furthermore, the demonstrated strain tunability of the spin-Hall anisotropy introduces an additional external degree of freedom for on-demand control of spin transport. These findings establish anisotropic ferromagnets with strong spin-orbit coupling as a versatile platform for exploring fundamental spin-orbit physics and for developing next-generation spintronic devices that leverage intrinsic material anisotropy for enhanced functionality and reduced energy consumption.

## Acknowledgments

The authors would like to acknowledge the financial support from National Key Research and Development Program of China (Grant No. 2022YFA1602701) and the National Natural Science Foundation of China (Grants No. 12574131, No. 12327806, and No. 12227806). Numerical calculation is supported by High-Performance Computing Center of Wuhan University of Science and Technology.